
\magnification=1200
\baselineskip=16pt plus 2pt
\hsize=15.5 true cm
\vsize=21. true cm
\def\lb{\hfil\break}
\def\to{\rightarrow}
\def\pslash{\mathord{\not\mathrel{p}}}
\def\kslash{\mathord{\not\mathrel{k}}}
\def\delslash{\mathord{\not\mathrel{\partial}}}
\def\zeroslash{\mathord{\not\mathrel{\hskip 0.04 cm 0}}}
\def\sqr#1#2{{\vcenter{\hrule height.#2pt\hbox
       {\vrule width.#2pt height#1pt\kern#1pt
       \vrule width.#2pt}\hrule height.#2pt}}}

\def\ddp{{d^dp \over {(2\pi)^d}}}
\def\ddk{{d^dk \over {(2\pi)^d}}}
\def\phivec{\vec\phi}
\def\Phivec{\vec\Phi}
\def\psvec{\vec\psi}
\def\psvecb{\vec{\bar\psi}}

\def\brezin{{[1]}}\def\ambj{{[2]}}
\def\zinn{{[4]}}
\def\damg{{[5]}}\def\neub{{[6]}}
\def\brez{{[7]}}\def\schn{{[8]}}
\def\abbo{{[9]}}\def\bm{{[10]}}
\def\gros{{[11]}}\def\mosh{{[12]}}
\hfill TECHNION-PH-54-92

\hfill November  1992
\vskip 0.5 true cm

\centerline {\bf On the Double Scaling Limit  of O(N)
Vector Models in d=2 }
\bigskip\bigskip  \smallskip
\centerline { Paolo Di Vecchia $^{\ddagger}$ }
\smallskip
\centerline{ \it Nordita, Blegdamsvej 17,
DK-2100 Copenhagen
$\zeroslash$ , Denmark}

\bigskip  \smallskip
\centerline { Moshe Moshe $^{* \ddagger}$}\smallskip
\centerline{ \it   Department of Physics, Technion -
Israel Institute of Technology}
\centerline{ \it   Haifa, ~~32000 ~~Israel}
\bigskip\bigskip
\vskip 2.0 true cm
\centerline {\bf ABSTRACT}

{ \narrower\narrower\smallskip
Recent interest in large N matrix models in the double scaling limit
raised  new interest also in O(N) vector models.  The limit
$N \rightarrow \infty$,  correlated with  the limit
 $g \rightarrow g_c$, results in an expansion in terms
of filamentary surfaces
and explicit calculations can be carried
out  also in dimensions $d\geq 2$.
It is shown  here that the
absence of physical massless bound states
in two dimensions sets strong constraints on this limit.}
\narrower
\vskip 2.0 true cm

\noindent
-----------------------------------------------------------------------------

\baselineskip=12pt plus 2pt
\noindent *~{\it Supported
in part by the GIF and the Henri Gutwirth fund}

\noindent $\ddagger$~{\it e-mail addresses : }
divecchia@nbivax.nbi.dk.bitnet

\hskip 2.7 cm phr74mm@technion.bitnet
\baselineskip=16pt plus 2pt

\vfill\eject

\baselineskip=16pt plus 2pt
Large N matrix models  providing representations of
dynamically triangulated random
surfaces in their double scaling
limit $^{\brezin}$ have recently
raised much interest.  Here, the surfaces are represented by
the  Feynman graphs  of the matrix  model in the limit, where
$N\to\infty$ and the coupling
constant $g \to g_c$ in a correlated manner.
An analogous interesting limit is the double scaling
limit in O(N) vector models, which represents
filamentary surfaces$^{\ambj-\zinn}$.
These models have many common features
with the matrix models and
give new insight into the nonperturbative structure associated
with the double scaling limit and  the  related underlying
hierarchy$^{\damg}$. In  many cases, the O(N)
vector models can be successfully studied$^{\zinn}$ also  in
dimensions $d\geq 2$ and thus gaining intuition for
the search of
a solution to  the long lasting
problem of understanding field theories in terms of
extended objects$^{\neub}$.
The double scaling limit in O(N) vector quantum field theories
reveals  an interesting new phase structure, as was
argued also in case of matrix models.

Consider the large N limit of the self
interacting scalar O(N)
symmetric vector model in d  Euclidian dimensions
defined by the functional integral
$$ Z_N=
\int {\cal D}\Phivec \, exp \bigl\{-
\int d^d x \,\{~ {1\over2}{(\partial_{\mu} \Phivec)}^2 +
U(\Phivec^2)~
\}~~ \bigr\} ~~$$
$$  =\int {\cal D}\Phivec\int {\cal D}\sigma\int
{\cal D}m^2 exp\bigl\{-\int d^d x \,
\{ {1\over2}{(\partial_{\mu} \Phivec)}^2 +
U(\sigma)-{m^2 \over 2}
(\sigma-\Phivec^2)\}~\bigr\} ~. \eqno[1]$$
$\Phivec$ is an N-component real scalar
field and the conventional treatment
of the large N limit  has been employed in Eq. [1].
The potential
$$U(\sigma)={\mu^2_0\over 2} \sigma +
{\lambda_0 \over 4}\sigma^2 \eqno[2]$$
will be discussed below.
After integrating  out $\tilde{\vec\phi}(x)~~$(where~$\vec\Phi(x)=
\vec\phi_c+\tilde{\vec\phi}(x)~)$ and
$\sigma(x)$, the  $m^2(x)$ integration
is carried out by a saddle point integration and the effective
action $S_{eff}\{\phi_c\}$
is obtained in the large N limit.
$L^{-d}S_{eff}\{\phi_c\}$
 is proportional to the free
energy per unit volume${^{\brez}}$  at fixed
$\vec\phi_c .$
One finds:
$$ \eqalign{ e^{ -S_{eff}
\{\phi_c\} } &=C e^{-{N L^d\over 2}\bigl\{ \int\ddp
\ln ~( p^2 + m^2 ) ~- {( 2 N
\lambda_0 )}^{-1} (m^2-\mu_0^2)^2 ~+
m^2\bigl( {\vec\phi_c^2\over N}\bigr)\bigr\}  }  \cr  &
\int {\cal D}\alpha(x)
e^{ -{N\over 2} \bigl\{  \int d^dx \alpha(x)
\bigl( \int\ddp
{1\over {p^2+m^2}} ~-~ {(  N \lambda_0 )}^{-1}
(m^2-\mu_0^2) ~+~\bigl ( {\vec\phi_c^2\over N}\bigr)  ~\bigr)
\bigr\}  } \cr & \hskip 1.5 cm  e^{ {N \over 4}
 \bigl\{ \int dx dy \alpha(x) \alpha(y)
\int \ddp e^{ip(x-y)} \bigl( \Sigma(p) + {1 \over
{N\lambda_0}} \bigr) ~~~+~~~ O(\alpha^3)
\bigr\}  }   }   \eqno[3]$$
where $m^2(x)=m^2+\alpha(x)$ and $m^2=m^2(\phivec^2_c)$
is the solution of the gap equation-saddle point condition:
$$ m^2 = \mu^2_0 + \lambda_0\biggl\{N\int
{d^dk \over (2\pi)^d} \bigl\{{1\over k^2 + m^2}\bigl\} ~~+~~
{\vec\phi_c}^2 \biggr  \} \eqno[4]$$

The double scaling limit$^{\ambj-\zinn}$ is reached in
the limit at which  $N \to \infty$
is correlated with the limit $\lambda \to \lambda_c$,
where $\lambda_c$
is the value of the coupling constant at which the
$O(\alpha^2)$ term at p=0 in Eq.[3] vanishes. Namely,
$$\Sigma(p=0) + {1 \over {N\lambda_0}} ~=~0~\eqno[5]$$
where
$$ \Sigma(p) = \int \ddk~
{1\over [(k-p)^2 + m^2]~[k^2 + m^2]}$$

In physical terms, the double scaling limit is reached when
the self coupling $\lambda_0$ of $\Phivec$
reaches the value $\lambda_c$ at which the strength
of the force between the quanta of the
$\Phivec$ fields, the fundamental
scalars  in the theory, binds them
together strongly enough to create
a {\bf massless} scalar, O(N)singlet,
$\Phivec\cdot\Phivec$ bound state.

Following the above procedure, it has been
shown in Ref.\schn ~that in
d=4 the effective potential is complex
and the critical vector model
cannot be consistently defined. This property is closely
related to the triviality of $\lambda \Phivec^4$ in d=4 in the
large N limit$^{\bm}$.   It is shown below that also in d=2 a
similar problem occurs and is associated
with the absence of massless bound states in two dimensions.

The divergences in Eq. [3] can be treated$^{\schn-\bm}$
in a conventional renormalization
scheme in the large N expansion
and the discussion in the last paragraph can be restated in
terms of renormalized physical parameters. In d=2, the gap
equation, Eq.[4], is thus given in terms of
the renormalized parameters by:

$$ m^2 - {\lambda N \over 4\pi}
\ln({M^2\over m^2}) = \mu^2 + \lambda
{\vec\phi_c}^2  \eqno[6]$$ where

$$\mu^2=\mu_0^2+{N\lambda\over 4\pi}\ln({\Lambda^2\over M^2})
\eqno[7]$$
and there is no renormalization of the coupling constant
$\lambda_0=\lambda$. (A momentum   U.V. cutoff has
been used in Eqs. [4] and [7].)
Since $\Sigma(0)= (4\pi m^2)^{-1}$,
the criticality condition in Eq. [5] implies
here :
$$ \lambda_c= -{4\pi m^2 \over N} \eqno[8]$$
which agrees with Eq. [3.23]  in the second paper of Ref. [4]
(note: $f={N\lambda\over 4}$).

The minimum of the effective action
can be read directly from Eq.[3],
namely (see also Refs. [8]-[9])
$$m^2\vec\phi_c=0~~~~~~~~~~~.\eqno[9]$$
It sets to zero the expectation value of the $\Phivec$ field in the
O(N) symmetric phase, where $m^2\neq 0$  is the physical mass of the
$\Phivec$ scalars.
The condition in Eq. [5] has to be satisfied, of course,
at the minimum of the effective potential, namely, at the values of
$m^2\neq 0$ and $\vec\phi_c =0$
for which  Eqs. [6] and [9] are satisfied.
The value of $m^2=m^2(\phivec^2_c =0) $ in
Eq.[8] is thus  the solution of
the gap equation (Eq. [6]) at $\phivec^2_c=0$, namely, at the minimum
of the effective action. In terms
of the renormalized parameters  and $\lambda=\lambda_c$
the gap equation is given  (at $\phivec^2_c=0$) by:

$$m^2\ln ({M^2e\over m^2})=\mu^2 \eqno[10]$$

The effective action in Eq. [3] in the
double scaling limit ($N \to \infty $ and
 $\lambda=\lambda_c$) is finally obtained by inserting there the
solution of the gap equation $m^2(\phi_c)$ found from Eq.[6].
In terms of the renormalized parameters,
Eq.[4] is given (at$\lambda=\lambda_c$
and $\mu^2$ from Eq.[10]) by:
$$ m^2(\phivec^2_c) + m^2 \ln ({m^2\over m^2(\phivec^2_c) e})=
-{4\pi m^2 \over N}\phivec^2_c  \eqno[11]$$

The right hand side of Eq.[11] has a
minimum at $ m^2(\phivec^2_c)=m^2$,
where it vanishes and thus there is no
real solution for $ m^2(\phivec^2_c)$ at
any real $\phivec^2_c\neq0$. Thus, the criticality
condition in Eq.[5] which implies a zero mass $\Phivec\cdot\Phivec$
bound state cannot be consistently defined.
The nonexistence of a real effective action
in the double scaling limit in d=2 is a reflection of
the absence of a zero mass bound state in two dimensions.
The absence of a consistent picture of the double scaling limit
in d=2  as a result of the vacuum fluctuations can be
clearly seen also in the self interacting spinor O(N) symmetric
theory defined by$^{\gros}$:
$$Z_N= \int{\cal D}\psvec{\cal D}\psvecb{\cal D}\sigma
e^{-\int d^dx[\psvecb
(i\delslash+g_0\sigma)\psvec -{1\over2}\sigma^2]}
\eqno[12]$$
After integrating out $\psvec$ and $\psvecb$,
the saddle point integration
on $\alpha(x)=\sigma(x)-\sigma_c$ results in a gap equation:
$$\sigma_c - N\int {d^2k\over (2\pi)^2}   Tr\bigl({
g_0\over \kslash+g_0\sigma_c}\bigr) =0
\eqno[13]$$
and a criticality condition (setting the
$O(\alpha(x)^2)$ term in the
functional integral to zero)
$$1 + g_0^2N\Sigma(p=0) = 0
\eqno[14]$$
where
$$ \Sigma(p)=\int{d^2k\over (2\pi)^2} Tr \bigl\{
{1\over (\kslash-\pslash+g_0\sigma_c)
(\kslash+g_0\sigma_c)}\bigr\}
\eqno[15]$$
As in the scalar case, Eq.[14] is proportional
to the four point  function
of the fundamental field, and the
criticality condition amounts to
imposing the existence of a massless
$\psvecb\cdot\psvec$ bound state in
the spectrum. No such pole exists in d=2, Eq.[14]
is never satisfied. One finds:
$$ 1 + g_0^2N\Sigma(p) ~=~ 1- g_0^2{N\over
2\pi}ln({\Lambda^2\over m^2})
  -g_0^2{N\over 2\pi}\sqrt{1+{4m^2\over p^2}}
\ln\biggl({ \sqrt{1+{4m^2\over p^2}}
-1\over \sqrt{1+{4m^2\over p^2}} +1}\biggr) \eqno[16]$$
Since $1- g_0^2{N\over 2\pi}ln({\Lambda^2\over m^2}) =
0$ as a result
of Eq.[13], the right hand side of Eq.[16] does not vanish
at $p=0$ and thus Eq.[14] cannot be satisfied.

Yet, another two dimensional model for
which the criticality condition
cannot be satisfied is the nonlinear
sigma model, where the equivalent
of Eq.[14] reads simply $\Sigma(p=0)=0$
and is clearly impossible to
satisfy.

In conclusion, the emerging physical picture  for
the double scaling limit
is very clear. In this limit, the force between the fundamental
quanta is set to the value at
which the  mass  of their bound state
vanishes. In d=2 dimensions, vacuum fluctuations avoid
reaching this limit. In the scalar theory the instability of
this limit is manifested by the absence of a real solution
to the relevant equations, namely,
the absence of a real effective action.
In the case of asymptotically free
theories the double scaling critical
condition is not satisfied for any
value of the coupling constant.
Thus, a formal analytic continuation is
required in order to give a
physical meaning to the double scaling
limit in d=2  dimensions.
In d=3 some of these problems are
absent and in certain cases
the massless bound state is associated
with the dilaton pole$^\mosh$.
\vskip 1 cm

\centerline{ ACKNOWLEDGMENTS}

M.M. thanks NORDITA for the hospitality
during  a visit last summer
and thanks  J. Feinberg for several useful discussions.

\vfill\eject
\bigskip
\centerline {\bf References}
\vskip 1 true cm

\def\PHL#1#2#3{Phys. Lett. {\bf#1B} #2 (19#3) }
\def\PRL#1#2#3{Phys. Rev. Lett. {\bf #1} #2 (19#3) }

\def\PRD#1#2#3{Phys. Rev. {\bf D#1} #2 (19#3)}
\def\NUP#1#2#3{Nucl. Phys. {\bf B#1} #2 (19#3)}

\item{1.} E. Br\'ezin and V.A. Kazakov , \PHL {236} {144} {90}
\item{}M.R. Douglas and S.H. Shenker,  \NUP {335} {635} {90}
\item{}D.J. Gross and A.A. Migdal, \PRL {64}{127}{90} ;
\NUP {340}{333}{90}
\item{2.} J. Ambjorn,  B.Durhuus and T. J\'onsson ,
\PHL {244}{403}{90}
\item{3.} S. Nishigaki and T. Yoneya , ~\NUP {348}{787}{91}
\item{} A.Anderson,  R.C. Myers and V. Periwal,
\PHL{254}{89}{91}   ~\NUP {360}{463}{91}.
\item{} P. Di Vecchia, M. Kato and N. Ohta,  ~\NUP  {357}{495}{91}.
\item{4.} J. Zinn-Justin, \PHL {257}{335}{91}
\item{} P. Di Vecchia, M. Kato and N. Ohta,  ~Int.  Journal of Mod.
Phys. {\bf 7A} 1391 (1992).
 \item{5.} P.H. Damgaard  and K. Shigemoto ~\NUP
{262}{432}{91} and references therein.
\item{6.}H. Neuberger, ~\NUP {340}{703}{90}
and private communications.
\item{7.}See e.g. E. Br\'ezin and
J. Zinn-Justin ~\NUP{257}{867}{85}
\item{} and aslo Refs. \schn-\bm .
\item{8.} H. J. Schnitzer Concerning the
Double Scaling Limit in the O(N)
Vector Model in Four Dimensions - Brandeis
preprint BRX-TH-333 (1992)
\item{} See also Ref. \abbo
\item{9.} L. F.  Abbott, J. S. Kang and
H. J. Schnitzer, ~\PRD{13}{2212}{76}
and references therein.
\item{10.}W.A. Bardeen and M. Moshe, ~\PRD
{28}{1372}{83}    \lb ~and~ \PRD {34}{1229}{86}.
See also Ref. \abbo
\item{11.} D. J. Gross and A. Neveu ~\PRD {10}{3235}{74}
\item{12.} M. Moshe - Proceedings of
Renormalization Group '91 Conference, Dubna, Sept. 1991, Ed. D. V.
Shirkov, World Scientific Pub. 1992.

\end